\documentclass[twocolumn]{aastex631}

\usepackage{amsmath}
\usepackage{bm}
\usepackage{booktabs}
\usepackage{longtable}
\usepackage{savesym}
\savesymbol{tablenum}
\usepackage{placeins}
\usepackage{siunitx}
\usepackage{threeparttable}
\restoresymbol{SIX}{tablenum}
\DeclareSIUnit{\parsec}{pc}
\DeclareSIUnit{\jansky}{Jy}
\DeclareSIUnit{\dmunit}{pc cm^{-3}}

\submitjournal{ApJL}

\shorttitle{The scattering timescale distribution function}
\shortauthors{Jow and Leung}
\graphicspath{{./}{figures/}}

\begin{document}

\title{FRB scattering statistics through the CGM are sensitive to morphology and intermittency}

\author[0000-0003-3236-8769]{Dylan L. Jow$^\ast$}
\affiliation{Kavli Institute for Particle Astrophysics \& Cosmology, Stanford University, Stanford, CA 94305, USA}
\email{dylanjow@stanford.edu}
\author[0000-0002-4209-7408]{Calvin Leung$^\ast$}
\affiliation{Department of Astronomy, University of California Berkeley, Berkeley, CA, 94720, USA}
\affiliation{Department of Physics, University of California Berkeley, Berkeley, CA, 94720, USA}
\affiliation{Miller Institute of Basic Research, University of California Berkeley, Berkeley, CA, 94720, USA}
\email{calvin\_leung@berkeley.edu}
\email{$^\ast$Equal contribution}



\begin{abstract}
The small-scale properties of circumgalactic gas in ordinary galaxies drive its bulk properties: the mass loading of cold neutral gas in galactic outflows affects their bulk momentum; gas cooling processes on small scales affect the spatial
distribution of gas in the cool ($T\sim 10^4\,{\rm K}$) circumgalactic medium (CGM). However, hydrodynamical
simulations have yet to resolve such small scales. Spectroscopy remains our primary probe of the small-scale CGM, with which sub-parsec scales are also challenging to resolve. Fast radio bursts (FRBs)---microsecond to millisecond duration radio pulses---are temporally
broadened (“scattered”) by gradients in the electron density transverse to the line of sight, often generated by fluctuations on the smallest spatial scales. This makes FRB scattering a complementary and scalable probe of the small-scale CGM.
We show that the distribution of scattering timescales induced by density fluctuations within a single,
foreground halo---the ``tau distribution function'' (TDF)---is sensitive to the
spatial morphology of the gas. The TDF is readily measurable and is analogous to areal covering factors reported in quasar absorption statistics. We compute the TDF in two regimes: scattering from a
turbulent, volume-filling medium (“volumetric scattering”); and scattering from discrete structures localized along the line of sight (“intermittent scattering”).  Within these regimes, the TDF is sensitive
to whether the cool gas comprises spherical, filamentary (1D), or sheet-like (2D) structures.
This work sets the stage for upcoming observations---from detectors like MeerKAT, Parkes, FAST, and the DSA---which will use hundreds of sight-lines through
nearby halos to probe the small-scale CGM.
\end{abstract}
\keywords{Radio transient sources (2008), Circumgalactic medium, Absorption spectroscopy}

\section{Introduction}\label{sec:intro}

The multi-phase nature of the circumgalactic medium (CGM) and ubiquitous presence of cool gas ($T \sim 10^4\,{\rm K}$) has been well-established by decades of observations~\citep{Rigby2002, Hennawi2006quasarsI, BowenCheclouch2011, tumlinson2017circumgalactic}. However, questions of the formation and survival of this gas remain unanswered. For example, the presence of cool gas in both star-forming and quiescent galaxies out to large radii has challenged standard pictures of cold accretion in which cool CGM clouds merely fuels star formation~\citep{FaucherGiguere2023}. Metals are found hundreds of kpc from the centers of their host galaxies --- farther than estimates from galactic winds would naively suggest~\citep{scannapieco2002early,veilleux2005galactic,peeples2014budget,chen2017mg}. 

``Cloudlet'' models in which cooling instabilities cause large clouds of warm gas to shatter into small-scale cloudlets of denser, cool gas are a class of models explaining these observations. 
They simultaneously explain the high covering factor of cool gas observed in quasar absorption spectra, as well as the survival of dense clouds in a hot medium~\citep{maller2004multiphase,mccourt2018characteristic,GronkeOh2020}. 
~\citet{mo1996gaseous} suggested that in order to survive the cool gas is distributed over a narrow range of small sizes, but cloud sizes remain uncertain and are difficult to directly resolve in realistic simulations~\citep{FaucherGiguere2023}.

Observations imply that the line of sight depth of cool clouds is $\sim 100$ pc~\citep{zahedy2021cubs}, but models have proposed even smaller characteristic scales (e.g.,~\citealp{mccourt2018,vedantham2019radio}; $\sim 0.1 - 10\,{\rm pc}$).

Recently, observations of non-thermal absorption-line broadening have permitted the non-thermal velocity dispersion of cool CGM gas to be measured~\citep{chen2024resolving,chen2024ensemble}. The aforementioned papers argue that since the non-thermal component of the velocity broadening is shared between all lines, it robustly probes cool gas velocities independent of photoionization modeling ~\citep{chen2026cgm}. At least one of the four systems of~\citet{chen2024ensemble} exhibits a velocity structure function (VSF) consistent with expectations for Kolmogorov turbulence -- a finding which is consistent with the prior expectation that cool clouds in the CGM should have high Reynolds numbers. While the origin of the turbulence is not yet clear, inferred energy transfer rates and energy dissipation timescales are consistent with both satellite interactions ``stirring'' the gas on tens of kpc scales and inflows from the IGM fragmenting on $\lesssim$ kpc scales. The VSFs of the remaining three systems are tentatively shallower, but are not constrained over a wide range of spatial scales, limited by seeing on leading ground-based facilities.

One way to probe CGM turbulence on transverse scales that are arbitrarily small is to leverage the scattering of fast radio bursts (FRBs;~\citealp[see e.g.][]{vedantham2019faraday, ocker2022radio, jow2024refractive, masribas2025refractive, ocker2025microphysics}). 
The pulse broadening of FRBs on $\tau \lesssim 100 \mu$s temporal scales correspond to angular deflections of $\alpha \sim \sqrt{c\tau / d_\mathrm{eff}} \sim \mu{\rm as}$ and may be generated by transverse electron density gradients on even finer spatial scales. 
Since density inhomogeneities in the CGM are dominated by the cool phase, FRB scattering potentially offers a complementary probe of the inner scale of a similar, if not the same, turbulent cascade probed on kiloparsec scales using spectroscopy. Finally, the large inferred volumetric rate of FRBs~\citep{ravi2019prevalence,shin2023inferring} is projected to translate into areal densities of $\gtrsim 1/$deg$^2$ in the next decade over the Northern sky~\citep{amiri2018chime,vanderlinde2019canadian,hallinan2019dsa}.

In this paper, we develop a novel observable that is sensitive to the spatial distribution and small-scale morphology of these electron inhomogeneities in the CGM on these unprecedented spatial scales. The ``scattering timescale distribution function'', or TDF, is the probability distribution of scattering timescales $\tau$ generated by the electron inhomogeneities within a halo. These scattering timescales are then imprinted on FRBs which intercept the halo at some effective distance $d_{\rm eff}$ within some impact parameter $b$. In practice, since there are several contributions to the observed scattering timescale, the measurement of a scattering timescale $\tau_{\rm sc}$ in observational data is an upper limit on any individual contribution\footnote{Put another way, the primary drawback of FRB scattering is that it is difficult to localize different contributions to the observed scattering along the line of sight for individual sightlines, though ensemble studies may eventually mitigate this shortcoming.}. Hence the observed TDF is expressed unambiguously only as a cumulative distribution $f(>\tau_{\rm sc})$, analogous to areal covering factors. In this form, $f(>\tau_{\rm sc})$ can be interpreted as the fraction of sight-lines through a halo with scattering timescales exceeding $\tau_{\rm sc}$.  

We make predictions for the TDF under the ``volumetric'' and ``intermittent'' theoretical formalisms for computing radio-wave scattering. Volumetric scattering is characterized by the continuous accumulation of small deflection angles as a ray traverses a volume filling medium. 
Much of the FRB scattering literature presumes that FRB scattering in the circumgalactic medium is volumetric \citep{cordes2016radio,ocker2022radio, faber2024heavily,Shin2025scattering, ocker2025microphysics}. 
For volumetric scattering by a turbulent medium, the scattering timescale for a given line of sight grows linearly with the effective path length intersecting the cool gas along that sight line. For a cloud complex of cool gas with size $R$ and a cool gas volume filling factor $f_V$, the mean intersection length through the cool gas is simply $f_V R$. 
However, the full distribution of intersection lengths, and thus the TDF, is dependent on whether the small-scale morphology of the gas, e.g., small spherical cloudlets, filamentary structures, or sheet-like structures (Sec.~\ref{sec:volumetric_scattering}). 

So-called ``intermittent'' scattering, due to spatially and/or temporally intermittent structures, may also arise \citep{jow2024refractive, masribas2025refractive}. In intermittent scattering as well as volumetric scattering, the TDF computed is also sensitive to the morphology of discrete structures (Sec.~\ref{sec:intermittent}). The consideration of intermittency in CGM scattering is motivated by their increasing relevance in models of the Galactic ISM on small scales. While the origin of FRB scattering has yet to be directly localized along the line of sight to discrete structures in the CGM (current $d_{\rm eff}$ measurements place the scatterers in either the host or Milky Way ISM, see e.g.~\citealp{masui2015dense,macquart2019spectral,schoen2021scintillation}), the greatly varying range of environments traversed by FRB sightlines suggests that it may be better-suited than a treatment which assumes a single power law. 

For instance, observations of pulsar scintillation show that intermittent structures dominate for at least a few Galactic sight-lines~\citep{brisken2010hundred, HRZhu2023, YHChen2025} and possibly in one FRB~\citep{kader2025detection}. Moreover, the framework for volumetric radio scattering in a turbulent medium commonly employed in the FRB literature assumes the electron density fluctuations are fully described by two-point Gaussian statistics (i.e. a power spectrum) between the inner- and outer-scale of the turbulence. This has, however, been challenged by modern understandings of turbulence, where intermittent structures and higher-order statistics play an important role \citep{Hopkins2013, Squire2017MNRASintermittency, Beattie2022}. 
Similarly, cosmic ray observations hint that intermittent, coherent structures in the ISM may be predominantly responsible for the scattering of low-energy (MeV - TeV) cosmic rays \citep{Butsky2024}; they may also play a role in resonant scattering of GeV cosmic rays~\citep{kempski2025unified}. 

\begin{figure}
    \centering
    \includegraphics[width=\linewidth]{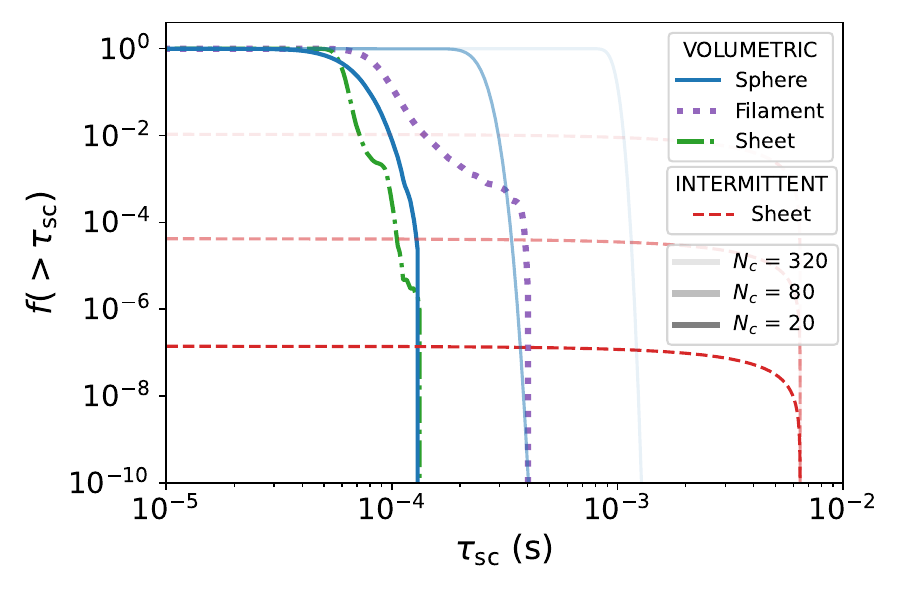}
    \caption{Predictions of the distribution of scattering timescale distribution function (the ``TDF''; equivalently the covering factor at a certain scattering timescale) for four different scattering scenarios. (Solid blue) Volumetric scattering due to a complex of size $R$ comprising spherical cool gas cloudlets of uniform radius, $r_c = 10\,{\rm pc}$. The different opacity curves are computed for differing cloudlet optical depths, $N_c = 20, 80$ and $320$, which correspond to $f_V R = 0.2, 0.8$, and $3.2\,{\rm kpc}$, respectively. We assume a mean-square-bending-angle-per-unit-length of $311\,{\rm \mu as^2 / pc}$ (see Section~\ref{sec:volumetric_scattering}). We place the scattering surface at a distance of $d_{\rm eff} = 0.8\,{\rm Mpc}$, corresponding to the distance to M31, and set the source wavelength to $\lambda = 75\,{\rm cm}$, corresponding to the low-end of the CHIME band. (Dotted purple / dot-dash green) An identical set-up to the blue curve with $N_c = 20$, except the cloudlets are given a filamentary / sheet-like geometry with an aspect ratio of $A = 500$, keeping the volume of the cloudlets fixed in all cases. (Dashed red) Intermittent refractive scattering from uniform sheets/slabs which have a density of $n_e = 10^{-2}\,{\rm cm}^{-3}$ and an aspect ratio of $A = 10^4$. The distance to the scatterer and wavelength are the same as in the previous cases, and we fix $f_V R = 0.2\,{\rm kpc}$. The opacity refers to the same values of $N_c$ as before, which correspond to sheet thicknesses of $\delta = 0.16, 0.04$ and $0.01\,{\rm pc}$, respectively.}
    \label{fig:summary}
\end{figure}

Fig.~\ref{fig:summary} summarizes the results of this paper by comparing predictions for the TDF under four different scenarios. In the sections that follow, we will describe how we obtain each of these curves in detail; however, for the moment, we will highlight some salient features of the results. First, we will note that the TDF for volumetric scattering in spherical cloudlets (the blue curve) follows a Gaussian distribution (i.e. an error-function when expressed as a CDF). Volumetric scattering in filamentary and sheet-like cloudlets (orange and green curves) deviates from Gaussianity: rays experience a bimodal distribution of path lengths through moderately-inclined and edge-on cloudlets, and, thus, the TDF exhibits a bimodality. The cumulative TDF for volumetric scattering in the aforementioned cases has an exponential drop-off toward large $\tau$, as the total path length through the scattering medium does not deviate far from the mean (see Section~\ref{sec:volumetric_scattering}). In contrast, the scattering time arising from refractive scattering due to intermittent and highly anisotropic structures with large aspect ratios increases smoothly with the inclination angle of the structure (up to some maximum value). Uniformly random inclination angles yield a TDF with a much shallower power-law drop off (see Section~\ref{sec:intermittent}). 


\section{Volumetric scattering in turbulent cool gas}
\label{sec:volumetric_scattering}

In this section, we will compute the TDF for complexes of cool gas comprising turbulent cloudlets of different geometries. A source that is volumetrically scattered by a turbulent medium with a power law spectrum of density fluctuations with sufficiently small inner scale attains a scattering timescale of (see Appendix~\ref{app:turbulent_scattering})
\begin{align}
\begin{split}
    \tau_{\rm sc} &= \frac{d_{\rm eff}}{2c}(1 + z_l)^3 L \eta, \\
    \eta &= A_\beta \lambda^4 r^2_e \widetilde{F}_l n^2_{e,c},
\label{eq:tau_volume}
\end{split}
\end{align}
where $d_{\rm eff} = d_{sl} d_{lo} / d_{so}$ is a combination of the distances between the source and scattering medium, the scattering medium and observer, and the source and observer. The redshift, $z_l$, is the redshift of the scattering medium, and $\lambda$ is taken to be in the frame of the scattering medium. The quantity $\eta$ is the mean square bending angle per unit length that rays undergo as they propagate through the medium. The background electron density of the cool gas is $n_{e,c}$. The fluctuation parameter $\widetilde{F}_l$ is an intrinsic gas property which in turn depends on the inner and outer scales of the fluctuations, and the variance of the electron density fluctuations evaluated at the outer scale (see Eq.~\ref{eq:def_fl}). The term $A_\beta$ is a numerical term that depends only on the index of the power-law spectrum, $\beta$. For Kolmogorov turbulence, $\beta = 11/3$ and the numerical constant is $A_{11/3} = \Gamma(7/6)$. 

Physically, Eq.~\ref{eq:tau_volume} implies that if the intrinsic properties of the density fluctuations are spatially uniform in the cool gas complex, then the TDF of that complex is entirely determined by the distribution of the effective path length, $L$, intersecting the cool gas, which we will hereafter refer to as the ``intersection length''\footnote{It is common in the scattering literature \citep[e.g.][]{Cordes2022redshift,cordes2016radio, ocker2022radio, Shin2025scattering}, to replace the factor of $d_{\rm eff} L n^2_{e,c} /2$ in Eq.~\ref{eq:tau_volume} with ${\rm DM}^2_l G_{\rm scatt}$, where ${\rm DM}_l = Ln_{e,c}$ is the excess dispersion measure from the scattering material and $G_{\rm scatt} = d_{\rm eff} / 2L$ is called the ``geometric leverage'' factor. The advantage of writing Eq.~\ref{eq:tau_volume} in this way is to relate the predicted scattering with a potentially observable quantity, ${\rm DM}_l$. However, for our purposes, the $F\times G$ formulation obscures the physical relationship between $\tau,~\eta$, and $L$.}. Acknowledging that these values vary substantially, we put some fiducial values to these expressions to get expressions for $\eta$ and $\tau_{\rm sc}$:
\begin{align}
\begin{split}
    \eta = 311 \, \frac{\rm \mu as^2}{\rm pc} & \, \frac{A_\beta} {\Gamma(7/6)} \left( \frac{\lambda}{75\, {\rm cm}} \right)^4 \\ 
    \times& \left( \frac{\widetilde{F}_l}{10^{-3}\,({\rm pc^2 \, km})^{-1/3}}\right) \left( \frac{n_{e,c}}{10^{-2}\,{\rm cm}^{-3}}\right)^2.
\end{split}
\end{align}
and
\begin{align}
    \begin{split}
        \tau_{\rm sc} = 0.12\,{\rm \mu s} \, (1 + z_l)^{-3}\left( \frac{d_{\rm eff}}{\rm Mpc} \right) \left( \frac{L}{\rm pc} \right) \left( \frac{\eta}{100\, {\rm \mu as}^2\,{\rm pc}^{-1}} \right).
    \end{split}
\end{align}
In what follows, we will adopt $d_{\rm eff} = 0.8\,{\rm Mpc}$, which corresponds to a source at infinity and a scattering surface at the distance to M31. We choose to focus on nearby halos as they have large extents on the sky and will have many FRB sight-lines through them, enabling constraints on their TDFs. Scattering through the M31 halo, specifically, will be the subject of forthcoming work. In addition, at a fixed observable scattering timescale $\tau_{\rm sc}$, nearby halos can be probed on the smallest transverse deflection scales $\alpha d_{\rm eff} \sim \sqrt{2 c \tau_{\rm sc} d_{\rm eff}}$.

\subsection{A single-scale mist}
\begin{figure}
    \centering
    \includegraphics[width=\columnwidth]{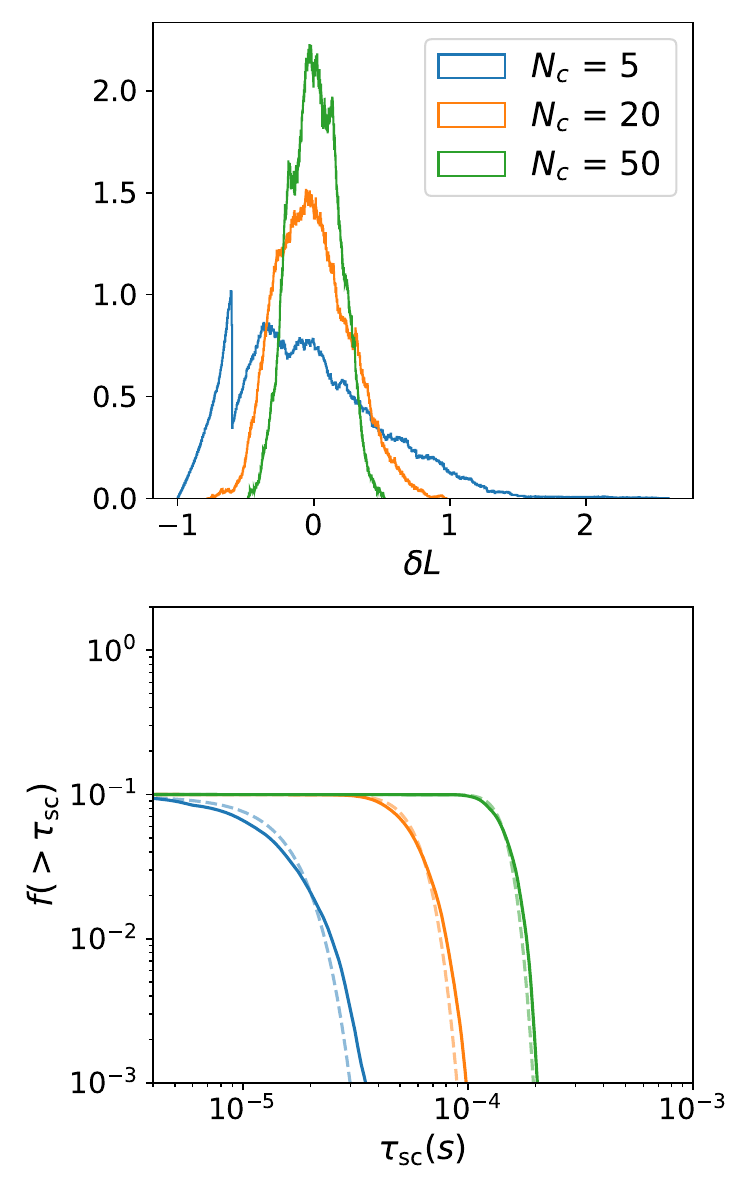}
    \caption{(Top) Distribution of intersection lengths for a complex of spherical cloudlets with radius $r_c = 10\,{\rm pc}$ and different cloudlet optical depths, $N_c = f_V R /r_c$. The distributions become narrower as $N_c$ increases, and shift horizontally with $\bar{L}$. (Bottom) The solid lines show the cumulative distribution functions for the scattering timescale (assuming $\eta = 311\,{\rm \mu as^2}/{\rm pc}$ and $d_{\rm eff} = 0.8\,{\rm Mpc}$) for the above intersection length distributions. The upper $x$-axis on the right panel shows the intersection length converted to a scattering time for $d_{\rm eff} = 0.8\,{\rm Mpc}$ and $\eta = 311\,{\rm \mu as}^2/{\rm pc}$. The dashed lines show the prediction of Eq.~\ref{eq:ftau_sphere_analytic}}
    \label{fig:mist_singlescale}
\end{figure}

We first consider a complex of cool gas of radius $R$ that comprises many smaller cool cloudlets with uniform radius, $r_c$. We will assume that these individual cloudlets have uniform electron densities, $n_{e}$, and fluctuation parameter, $\widetilde{F}_l$, so that each cloudlet bends rays by the same angle squared per unit distance, $\eta$. Thus, the statistics of the scattering time for sight-lines through this complex of cool gas is given by the statistics of the intersection length, $L$, through the individual cloudlets. Now, let the cool cloudlets have a volume filling fraction $f_V$ of the total volume of the complex. The total number of cloudlets in the complex is $f_V R^3 /r_c^3$, and the average path length through the cloudlets intersected by a line of sight is $\overline{L} = f_V R$. The average \textit{number} of cloudlets intersected by a sight line (the cloudlet optical depth) is given by
\begin{equation}
    N_c= f_V R / r_c.
    \label{eq:n_c}
\end{equation}
Fig.~\ref{fig:mist_singlescale} shows the distribution of $L$ for sight lines through a region of length $R = 1\,{\rm kpc}$ filled with spherical cloudlets of radius $r_c = 10\,{\rm pc}$, for different volume filling factors, $f_V$. We compute these distributions numerically by populating a box of dimensions $R \times 20r_c \times 20r_c$ with a number, $f_V R^3 / r_c^3$, of cloudlets with uniform random co-ordinates. A line of sight with a transverse displacement of $b$ relative to the centre of a cloudlet has an intersection length through the cloudlet given by
\begin{equation}
    L_c = 
    \begin{cases}
        2 \sqrt{r^2_c - b^2} & \text{if } b \leq r_c,\\
        0  & \text{if } b > r_c.
    \end{cases}
    \label{eq:Lc_sphere}
\end{equation}
For a fine grid of sight-lines through the cloudlet complex, we sum the intersection lengths through every cloudlet that is intersected by the sight line. The resulting distribution is approximately Gaussian with $L/r_c \sim \mathcal{N}(N_c, N_c)$ for $N_c > 1$. Alternatively, we can define
\begin{align}
    \delta L = \frac{L - \overline{L}}{\overline{L}} = \frac{L}{f_V R} - 1,
\end{align}
which is distributed according to $\delta L \sim \mathcal{N}(0,N_c^{-1})$. Defining $\overline{\tau}$ as the mean scattering time due to an intersection length of gas $\overline{L} = f_V R$, the corresponding TDF is
\begin{align}
\begin{split}
    &f(>\tau) = \frac{1}{2} {\rm erfc} \left(\frac{\frac{\tau}{\overline{\tau}} - 1}{\sqrt{2 / N_c}} \right), \\
    \label{eq:ftau_sphere_analytic}
\end{split}
\end{align}
The bottom panel of Fig.~\ref{fig:mist_singlescale} compares the empirical curves (solid) and the approximate curve (dashed) computed using Eq.~\ref{eq:ftau_sphere_analytic}. The agreement improves for larger $N_c$, but in either case the qualitative picture is clear: turbulent scattering due to mists produces an exponential cutoff in the $f_{\rm sc}(>\tau)$ curve above $\tau =\overline{\tau}$. Note that there are two effective degrees of freedom: the distribution of intersection lengths, and thus the TDF, only depends on the complex size, $R$, and the volume filling factor, $f_V$, through the combination $f_V R$: whereas $f_V R$ sets the position of the cutoff, $N_c$ sets it sharpness. 

\subsection{Sheets and filaments versus cloudlets}

In arguing that the statistics of the scattering timescale are highly dependent on the statistics of the intersection length through the cool gas, it becomes clear that the scattering timescale statistics are sensitive to the geometry of the cool gas. Here we will compare three simple scenarios as illustrative of this point: gas comprising cloudlets in the shapes of spheres, filaments, and sheets. The intersection length through a filament or sheet is highly dependent on the inclination angle of the object relative to the line of sight, unlike a sphere. Thus, the statistics of the intersection lengths through filaments and sheets will deviate from the simple Gaussian distribution that we have already shown characterizes spherical cloudlets. 

\begin{figure}
    \centering
    \includegraphics[width=\linewidth]{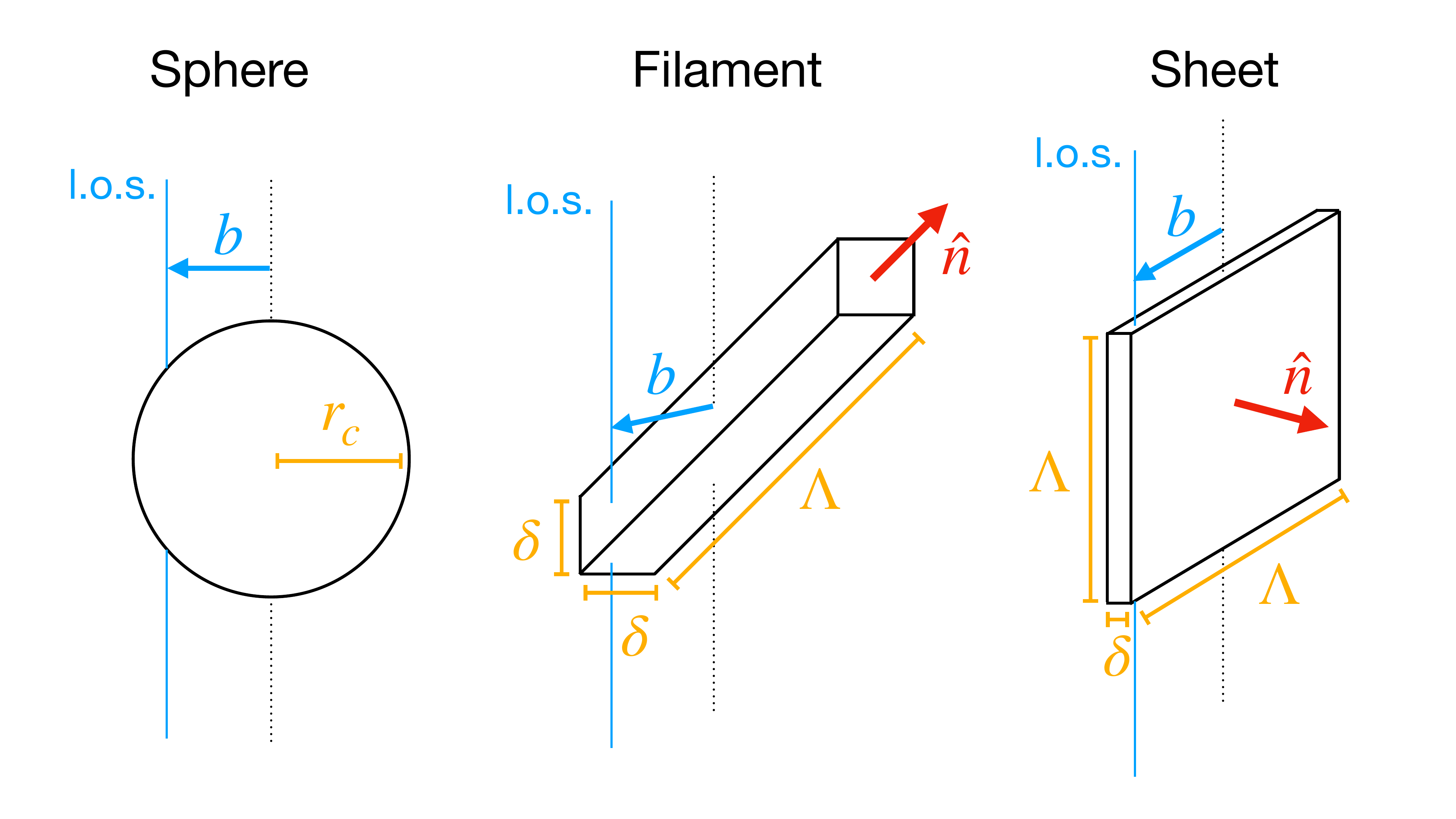}
    \caption{Diagram showing the geometric parameters used in Eqs.~\ref{eq:Lc_sphere}, \ref{eq:Lc_filament}, and \ref{eq:Lc_sheet} to define the intersection length through a spherical, filamentary, or sheet-like cloudlet, for a line of sight with impact parameter ${\bf b}$ from the centre of the cloudlet.}
    \label{fig:sfs_diagram}
\end{figure}

For a sphere of radius $r_c$, Eq.~\ref{eq:Lc_sphere} gives the intersection length through the sphere at an impact parameter of $b$ from the centre. Consider now a rectangular filament of dimension $\delta \times \delta \times \Lambda$. Let $\hat{n} = (\sin\theta\cos\phi, \sin\theta\sin\phi, \cos\theta)$ be the vector pointing along the long edge of the filament, and let $\theta$ be the inclination angle relative to the line of sight (see Fig.~\ref{fig:sfs_diagram} for a diagram showing the geometric parameters). We take the intersection length through a filament centred at the origin, intersected by the line of sight to be
\begin{equation}
    L_c^{\rm filament}({\bm b}) = 
    \begin{cases}
        {\rm min} \left\{ \frac{\delta}{|\sin\theta|}, \Lambda\right\}, &\tilde{b}_y < \delta\,\,{\rm and}\,\, \\ &\tilde{b}_x < {\rm max}\left\{ \frac{\Lambda}{2} |\sin \theta|, \frac{\delta}{2} \right\} \\
        0  & \text{else},
    \end{cases}
    \label{eq:Lc_filament}
\end{equation}
where $\tilde{b}_x = b_x \cos\phi + b_y \sin \phi$ and $\tilde{b}_y = - b_x \sin\phi + b_y \cos\phi$. The conditions on $\tilde{b}_x, \tilde{b}_y$ ensure that the line of sight passes through the filament. When the inclination angle is shallow, the intersection length is just the thickness $\delta / |\sin\theta|$. However, when the inclination angle is sufficiently large, the line of sight passes through the entire length of the filament, $\Lambda$. Since $\Lambda$ sets an upper bound, we take the minimum of these two values for simplicity. 

Now consider a sheet of dimension $\delta \times \Lambda \times \Lambda$ with $\delta \ll \Lambda$. Let $\hat{n}$ be the vector normal to the face of the sheet (see Fig.~\ref{fig:sfs_diagram}). The intersection length for a sheet centred at the origin is
\begin{equation}
    L_c^{\rm sheet}({\bm b}) = 
    \begin{cases}
        {\rm min} \left\{ \frac{\delta}{|\hat{n} \cdot \hat{z}|}, \Lambda\right\}, &\tilde{b}_y < \delta\,\,{\rm and}\,\, \\ &\tilde{b}_x < {\rm max}\left\{ \frac{\Lambda}{2} |\hat{n} \cdot \hat{z}|, \frac{\delta}{2} \right\} \\
        0  & \text{else}.
    \end{cases}
    \label{eq:Lc_sheet}
\end{equation}
We now compare the length distribution for different cloud morphologies, all with a fixed $\bar{L} = f_V R = 0.2\,{\rm kpc}$. For the spherical clouds, we choose a radius of $r_c = 10\,{\rm pc}$. We fix the volume to be the same for the filaments and sheets, such that for the filaments $\frac{4}{3} \pi r^3_c = \delta^2 \Lambda$ and for the sheets $\frac{4}{3} \pi r^3_c = \delta \Lambda^2$. Thus, the gas complex has the same number of objects in each case, which we randomly distribute throughout the total volume. This leaves the choice of the aspect ratio, which we fix to $\Lambda / \delta = 500$ for both the filaments and sheets. 

\begin{figure*}
    \centering
    \includegraphics[width=2\columnwidth]{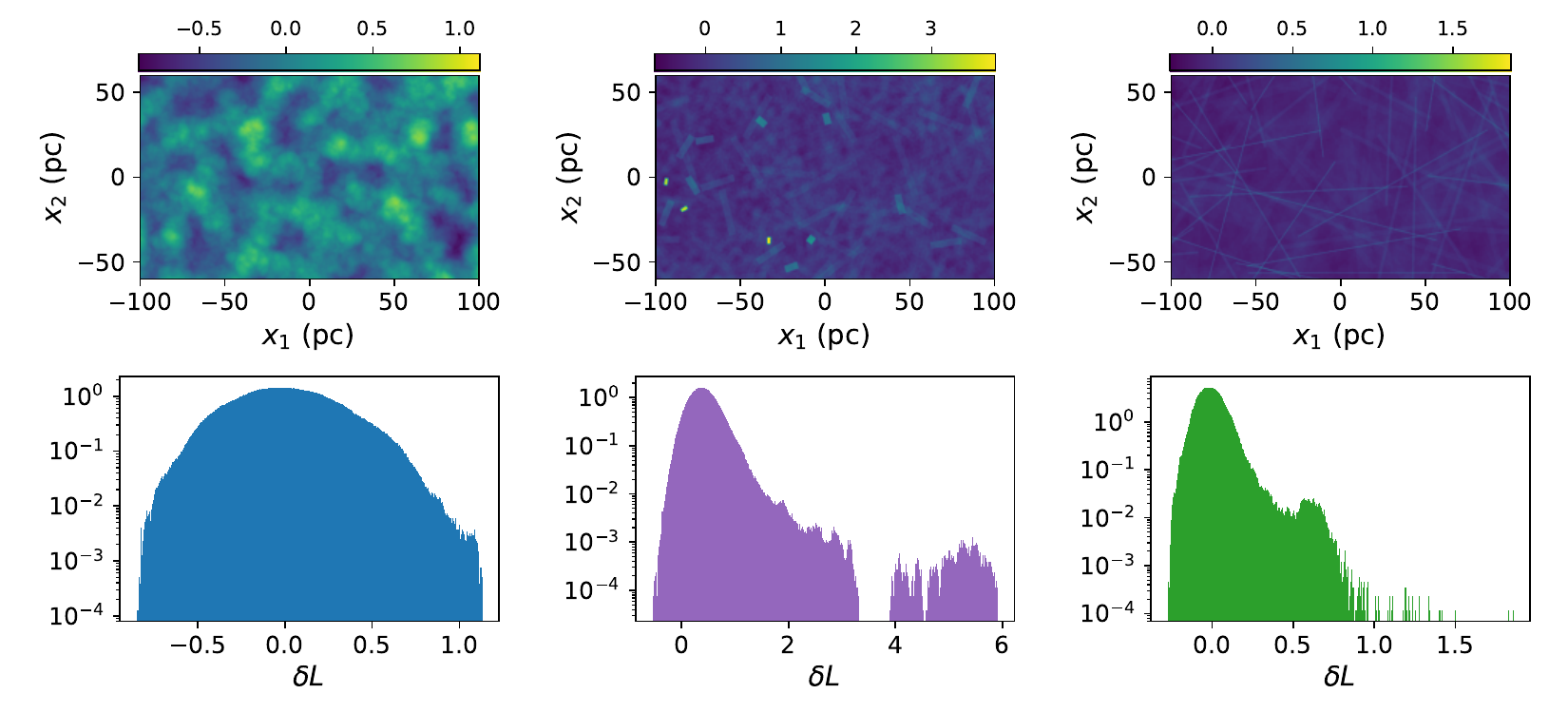}
    \caption{(Top row) The intersection lengths of a mist of spherical (left), filamentary (middle), and sheet-like (right) cloudlets through a box of dimension $0.2 \,{\rm kpc} \times 0.2 \,{\rm kpc} \times 1\,{\rm kpc}$. The spheres have uniform radius $r_c = 10\,{\rm pc}$. The filaments have dimension $\delta \times \delta \times \Lambda$ and the sheets have dimension $\delta \times \Lambda \times \Lambda$, where in both cases we choose $\Lambda / \delta = 500$. The smaller dimension is chosen so that the volume of each object is equal to the $\frac{4}{3}\pi r_c^3$. Each box is randomly populated with cloudlets to have a volume filling factor of $f_V = 0.2$. The colour map shows the intersection length contrast $\delta L = (L -\overline{L}) / \overline{L}$ where $\overline{L} = f_V R$.  (Bottom row) The distribution of the intersection length contrast values in the top row.}
    \label{fig:cfs}
\end{figure*}

\begin{figure*}
    \centering
    \includegraphics[width=2\columnwidth]{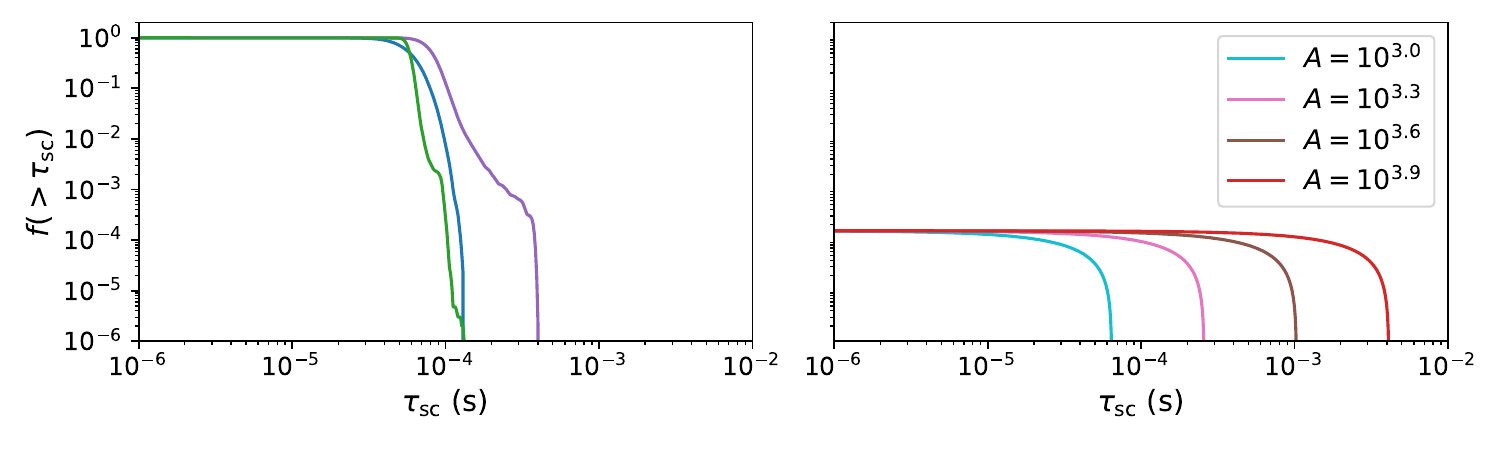}
    \caption{(Left) The cumulative TDF for volumetric scattering through complexes with spherical, filamentary, and sheet-like cool gas clumps (derived from the intersection length distributions shown in Fig.~\ref{fig:cfs}). The intersection lengths are converted to scattering times with $\eta = 311\,{\rm \mu as}^2/{\rm pc}$ and $d_{\rm eff} = 0.8\,{\rm Mpc}.$ (Right) The cumulative TDF for itermittent scattering from a complex of refractive slabs with $N_c = 200$, $f_VR = 0.2\,{\rm kpc}$, and varying aspect ratios. We assume the slabs have uniform density $n_e = 10^{-2}\,{\rm cm}^{-3}$ and are at a distance of $d = 0.8\,{\rm Mpc}$. The $y$-axis is the same as the left plot.}
    \label{fig:tdf_cdfs}
\end{figure*}

The $\delta L$ distributions are shown in Fig.~\ref{fig:cfs}. The top row shows the intersection length projected onto the sky. The qualitative difference between spheres, filaments, and sheets is readily apparent. The largest values for the filament case are dominated by roughly square regions where the filament is parallel to the line of sight. The largest values for the sheet case are dominated by thin streaks representing the edge of the sheet when the sheet is parallel to the line of sight. The bottom row shows the distribution of the deviations of the intersection lengths from the mean, $\delta L$. In all cases, the mean intersection length is roughly given by $\overline{L} = f_V R$ as expected. The left panel of Fig.~\ref{fig:tdf_cdfs} compares the resulting TDF. All three distributions share the same mean, $\overline{\tau}$, corresponding to the mean intersection length; however, the extended tails of the filament and sheet distributions cause the TDF to deviate from the simple Gaussian expectation. 

\section{Intermittent refractive scattering}
\label{sec:intermittent}

In the previous section, we considered volumetric scattering through a statistically isotropic medium with turbulent density fluctuations. In that setting, the scattering statistics are determined by the total length through which a given line of sight intersects the scattering medium. Different gas morphologies give rise to distinguishable statistics in the observed scattering timescales. However, this framework relies on a few key assumptions. First, it assumes that the scattering is volumetric; i.e., rays accrue small deflections continuously as they propagate through the medium. Second, it assumes that the medium responsible for scattering is statistically isotropic and characterized by a power spectrum of density fluctuations. This is in contrast with \textit{intermittent} scattering, where the scattering angle is dominated by a small number of coherent structures along the line of sight. Both of these assumptions have been challenged in different contexts. Recent studies of turbulence emphasize the importance of spontaneous formation of intermittent structures in turbulent media \citep{Hopkins2013, Squire2017MNRASintermittency, Beattie2022}. Modeling a turbulent medium as a Gaussian random field with a power-law power spectrum is likely to miss important effects even for standard Kolmogorov turbulence. Moreover, detailed study of a few nearby pulsars has shown that scattering through at least some lines of sight through the ISM is dominated by intermittent structures\citep{brisken2010hundred, HRZhu2023, YHChen2025}. Similarly, cosmic ray observations paint a similar picture: intermittent, coherent structures in the ISM may be predominantly responsible for the scattering of low-energy (MeV - TeV) cosmic rays \citep{Butsky2024}. Thus, both theoretical considerations and observational results across multiple domains point to the increasing relevance of intermittency to models of the ISM. The CGM, of course, is a very different context from the ISM. However, in the absence of observations indicating one way or another, when making predictions for radio scattering in the CGM, we should not dismiss the possibility that intermittent structures may play an important role. Thus, in this section, we consider possible sources of intermittent scattering in the CGM.

\subsection{Refractive scattering from smooth, spherical clouds}

As \cite{jow2024refractive} argue, even if the inner scale of the turbulence is too large to produce significant volumetric scattering, cool gas cloudlets can still produce significant large-scale refractive scattering if the electron densities are sufficiently high. In that scenario, each cloudlet acts as a single, smooth plasma lens. If the bending angle is sufficiently large, an ensemble of such cloudlets can produce many images that will be observed as a scattering tail. 

A light ray propagating through a plasma accrues a bending angle of 
\begin{equation}
    \alpha = (1 + z_l)^{-2}\frac{\lambda^2 r_e}{2\pi} \nabla_\perp N_e,
\end{equation}
where $\nabla_\perp N_e$ refers to the gradient of the electron column density perpendicular to the line of sight. The wavelength, $\lambda$, is the observing wavelength, and so the bending angle has a factor of $(1+z_l)^{-2}$, where $z_l$ is the redshift of the scattering surface. 

\subsection{Refractive scattering from smooth, spherical clouds}
For a spherical cloudlet of electrons, the column density is related to the 3D density by $N_e = r_c n_e$, where $r_c$ is the size of the cloudlet. For a smoother, spherical cloudlet, the column density gradient can be estimated by $\nabla_\perp N_e \sim N_e / r_c \sim n_e$. Thus, the bending power of a spherical cloudlet is roughly
\begin{equation}
    \alpha_{\rm sphere} \sim (1+z_l)^{-2}\frac{\lambda^2 r_e n_e}{2\pi}.
    \label{eq:alpha_sphere}
\end{equation}
We will show that for fiducial parameters for the cool CGM, a single spherical cloud does not have sufficient bending power to produce multiple refractive images, i.e. that the convergence $\kappa < 1$. In order to form multiple images, more than one ray needs to be bent into the observer's line of sight. The maximum transverse distance to the un-perturbed line of sight that a lens can be placed while still bending rays that intersect the observer is $x = \alpha d_{\rm eff}$. The criterion for a lens to form multiple images is that this distance be greater than the transverse size of the lens itself: $x > r_c$. The ratio $x / r_c$ is precisely $\kappa$~\citep[see][]{jow2024refractive}, and for fiducial parameters is given by 
\begin{align}
\begin{split}
    \kappa_{\rm sphere} \sim 0.03 \,(1+ z_l)^{-2} &\left(\frac{\lambda}{75\,{\rm cm}} \right)^2
    \left( \frac{d_{\rm eff}}{1\,{\rm Gpc}} \right)\\
    \times &\left(\frac{n_e}{10^{-2}\,{\rm cm}^3}\right) \left( \frac{r_c}{10\,{\rm pc}} \right)^{-1}.
\end{split}
\end{align}
Therefore, if the cool gas comprises spherical cloudlets with $r_c \sim 10\,{\rm pc}$, we do not expect significant refractive scattering from individual cloudlets, consistent with \citet{masribas2025refractive}. The length scale of density fluctuations may be too large (i.e. gradients are too small), or the electron densities too small, to produce significant volumetric scattering while remaining in pressure equilibrium with the hot CGM. 

One way that spherical cool gas cloudlets can still produce significant large-scale refractive scattering if collections of individual scatterers are coherently organized~\citep{jow2024refractive}, e.g. into large-scale sheets~\citealp{pen2014pulsar}. In this scenario, each cloudlet acts as a single, smooth plasma lens, and the ensemble of such cloudlets, once projected onto the lens plane, can produce many images that will be observed as a scattering tail. 

An ensemble of such cloudlets has a collective convergence of $\kappa_{\rm tot} \sim \sqrt{N_c} \kappa_{\rm sphere}$, where $N_c = f_V R / r_c$ is the average number of cloudlets that intersect a given line of sight. In order for an ensemble of such cloudlets to produce many refractive images, hundreds of cloudlets would need to be intersected by the average line of sight. Note, however, that the total number of refractive images that are formed is roughly $\sim \kappa_{\rm tot}^2$. This scales only linearly with the volume filling factor, but is highly sensitive to the cloud size at fixed $f_V$, scaling as $\sim r_c^{-3}$. Thus, even if only a small fraction of the total volume of cloudlets comprises sub-parsec cloudlets, intermittent refractive scattering is still possible from spherical cloudlets.

\subsection{Intermittent scattering by smooth, refractive slabs}

In this section, we consider intermittent, refractive scattering from uniform sheets/slabs of free electrons. High aspect-ratio sheet structures have been proposed for the morphology of the warm neutral interstellar medium~\citep{heiles2003millennium}. Elongated sheet structures that are inclined along the line of sight have also been invoked to alleviate the so-called ``over-pressure problem'' which arises from the assumption that the scattering structures are spherical.
Briefly, $\sim {\rm mas}$ bending angles that are routinely observed in pulsar scintillation, if generated by spherical small-scale scattering structures, yield measurements of the electron density (Eq.~\ref{eq:alpha_sphere}) that are three-orders of magnitude over-pressured relative to the ambient ISM. Anisotropic scatterers -- slabs with large ($\sim 10^3$) aspect ratios -- alleviate this problem, since the bending angle is enhanced by a factor of the aspect ratio, $A$ (see Appendix~\ref{app:slab}). In the CGM, magnetic fields may act to form sheet-like structures. \citet{mccourt2018} originally proposed that slightly overdense gas in the CGM shatters into tiny spherical cloudlets as it cools, with a characteristic radius set by the sound speed times the cooling time. Later, hydrodynamical simulations confirmed this picture. However, magnetic fields likely play an important role in the gas dynamics of the CGM, and will induce anisotropy in the cool gas structures. \cite{wangoh2025} find that magnetic pressure prevents shattering into spherical clouds. Instead, gas is confined to flow parallel to the magnetic fields, and counter-streaming motions produce elongated slabs of enhanced electron density. While the physics of the CGM is uncertain, highly anisotropic sheets (with large aspect ratios) have been seen to be produced in the presence of magnetic fields across multiple contexts and we are motivated to study their scattering properties.

In Appendix~\ref{app:slab}, we introduce a simple model for refractive scattering by uniform clouds of electron density confined along one dimension with large aspect ratios, which we refer to as ``slabs". An individual slab of electrons with density $n_e$ and aspect ratio $A$ produces a bending angle as a function of the slab's inclination $\iota$ relative to the line of sight:
\begin{equation}
    \alpha(\iota) = (1 + z_l)^{-2}\frac{\lambda^2 r_e n_e}{2 \pi} \min\left\{ \tan \iota, A \right\}. 
    \label{eq:alpha_sheet}
\end{equation} 
The lens convergence $\kappa$ of such slabs is also enhanced by a factor of $A$ relative to a spherical cloud. The scattering time as a function of inclination angle for a single sheet is $\tau_{\rm corr}(\iota) = d_{\rm eff} \alpha_{\rm sheet}(\iota)^2 / 2c$, where we take the characteristic bending angle $\alpha_{\rm slab}(\iota)$ to be half the maximum bending angle. Consider a complex of gas of size $R$ filled with refractive slabs with uniform thickness $\delta$. In Appendix~\ref{app:slab}, we find that the average number of slabs intersected by the line of sight through such a complex is given by
\begin{equation}
    N_c = \frac{1}{2 \pi \left(1 + \log A\right)} \frac{f_V R}{\delta}.
    \label{eq:Nc_corr}
\end{equation}
In Appendix~\ref{app:slab}, we also compute the probability that an ensemble of such slabs refracts rays into the line of sight above some bending angle. This is just the TDF and is given by:
\begin{align}
\begin{split}
    &f(>\tau) = 1 - \exp\Big\{ \\
    &-\pi \frac{f_V R}{A^2 \delta^3} d_{lo}^2 \alpha_0^2
    \Big(\left[ x \sqrt{x^2 + 1} - \arctan \left( \frac{x}{\sqrt{x^2+1}}\right) \right]^{x=A}_{x=\alpha/\alpha_0} \\
    & + \frac{A^2}{\sqrt{A^2+1}}\Big) 
    \Big\} 
    \label{eq:TDF_slab}
\end{split}
\end{align}
where $\alpha_0 = \lambda^2 r_e n_e / 4 \pi$.

The right panel of Fig.~\ref{fig:tdf_cdfs} shows $f(>\tau)$ for a complex of cool gas filled with slabs of varying aspect ratio such that $N_c = 200$. For a complex of size $R = 1\,{\rm kpc}$, a volume filling factor of $f_V = 0.05$, and an aspect ratio of $A = 1000$, this corresponds to a thickness of $\delta = 0.02\,{\rm pc}$. The curves are identical for different aspect ratios up until they hit the maximum scattering time, which is set by
\begin{align}
\begin{split}
    \tau_{\rm max} &= \frac{d_{\rm eff}}{2 c} (1 + z_l)^{-3} \alpha^2_0 A^2, \\
    &= 0.08\,{\rm s} \, (1+z_l)^{-3} \,
    \left( \frac{d_{\rm eff}}{1\,{\rm Gpc}} \right) \left(\frac{\lambda}{75\,{\rm cm}} \right)^4 \\
    &\,\,\,\,\,\,\,\,\,\,\,\,\,\,\,\,\,\,\,\,
    \,\,\,\,\,\,\,\,\,\,\,\,\,\,\,\,\,\,\,\,\,\,\,\times \left(\frac{n_e}{10^{-2}\,{\rm cm}^3}\right)^2 \left( \frac{A}{10^3} \right)^2.
\end{split}
\end{align}
The $f(>\tau)$ curve is largely flat up until this point; however, unlike the curves for the volumetric scattering, they do not approach unity as $\tau \to 0$. Whereas volumetric scattering produces a continuous range of bending angles out to some characteristic angle, a refractive lens with bending angle $\alpha$ only scatters light into the line of sight if it is at an impact parameter $d_{\rm eff} \alpha$ in the scattering plane. Thus, there is only a finite probability that any scattering is produced at all (see Appendix~\ref{app:slab}):
\begin{equation}
    f(>0) = 1 - \exp \left\{ -  \pi \frac{f_V R d_{lo}^2}{\delta^3} \alpha^2_0 \right\}.
\end{equation}
Here we have focused on intermittent scattering from a collection of slabs with some fixed aspect ratio. We might instead consider the intermittent scattering from a collection of filaments with random orientations. However, the probability that a filament is highly inclined relative to the line of sight is suppressed relative to the probability for a sheet/slab. Namely, if $\hat{n}$ defines the orientation of the filament and lies parallel to the filament (see Fig.~\ref{fig:sfs_diagram}). In that case, the bending angle increases monotonically with \textit{decreasing} $\iota$ and it is less likely for a filament to be nearly aligned with the line of sight than it is for a sheet. In the former case, $\hat{n}$ must be close to the pole, and, in the latter case, $\hat{n}$ only need lie near the equator. Nevertheless, it is straightforward to amend the derivation in Appendix~\ref{app:slab} for filaments.

\section{Discussion}
\label{sec:discussion}

In this paper, we have examined different possible scenarios for FRB scattering by inhomogeneous small-scale structures in the CGM. 
In both the volumetric and intermittent scattering formalisms, we show by explicit construction of spherical, filamentary, and sheet-like plasma substructures how the probability distribution of scattering timescales -- the TDF -- is sensitive to the small-scale geometry and isotropy of the turbulent gas. 
Spherical cloudlets have been shown to form when over-densities cool faster than their sound crossing time, shattering into a mist of spherical cloudlets \citep{maller2004multiphase,mccourt2018, GronkeOh2020}. In the presence of magnetic fields, however, these shattered cloudlets can become highly anisotropic with large aspect ratios \citep{wangoh2025}. 
Filamentary structures may naturally form when spherical cloudlets become entrained in the hot gas, forming cometary tails as they move within the hot phase \citep{McCourt2015hotwinds}. 
Thus, determining the morphology of the cool gas through FRB scattering may directly implicate the physical processes at work in the CGM at small scales. 

The statistics of volumetric scattering are determined by the total path length through the scattering material. In contrast, intermittent scattering is likely to be dominated by highly anisotropic structures with large aspect ratios, as these are precisely the structures that can produce large column density gradients, without excessive intrinsic densities. Therefore, the statistics of intermittent scattering are determined by the random orientations of the structures relative to the line of sight. Whether scattering in the CGM will be dominated by volumetric scattering or intermittent scattering is unknown. However, it is worth noting that pulsar scintillation observations have revealed the presence of intermittent refractive scattering sites in the ISM for some sight-lines~\citep{HRZhu2023, YHChen2025}. Similarly, recent observations of cosmic rays have suggested that low-energy cosmic rays are also scattered not volumetrically in the ISM but by a small number of coherent and intermittent structures: potentially the very same structures that are responsible for pulsar scintillation and extreme scattering events~\citep{kempski2025unified}. Moreover, while the volumetric scattering formalism that has been used here is commonly employed in the FRB scattering literature~\cite[e.g.][]{cordes2016radio,ocker2021constraining}, Eq.~\ref{eq:tau_volume} is ultimately derived assuming that the underlying turbulent medium is well-characterized by a power spectrum (i.e. its two-point statistics) (see Appendix~\ref{app:turbulent_scattering}). This simple picture of turbulence is inconsistent with modern understandings of turbulence. The so-called ``intermittency modeling" paradigm of turbulence studies increasingly emphasizes the importance of spontaneous, coherent and intermittent structure in the morphology and dynamics of turbulent media. Bridging the gap between FRB scattering observations and the physics of the CGM will require moving beyond simple scattering prescriptions. 

By introducing the tau distribution function (TDF), this work provides a conceptually simple framework that may be applied to observations to begin to probe the microphysics of the CGM. However, since there are many potential sources of scattering outside of intervening CGM halos (including the Milky Way and host ISM), and since, in practice, scattering often falls below detectability thresholds, there remain practical challenges in directly measuring the TDF for a given halo. By taking the total amount of observed scattering from all contributions as an upper limit on the amount of scattering that can be contributed by a halo, the TDF may observationally be constrained (Leung et al., in prep). A significant limitation to this approach arises from the unknown intrinsic TDF of the FRB host environment. Localization of sources may enable the removal of FRBs in environments which are likely to significantly dominate the observed scattering (such as FRBs in edge-on galaxies or active star forming regions). Alternatively, selecting FRBs without any foreground halos may enable a measurement of the intrinsic TDF, which can then be accounted for when placing constraints on the TDF of intervening halos. Further exploration of these practical issues will be the subject of forthcoming work.

\section{Conclusion}
In this paper, we have proposed to measure the scattering timescale distribution function, or TDF, of intervening galaxy halos to characterize their CGM. We have revisited the volumetric scattering formalism originally parameterized by \citet{cordes2016radio}, and have shown that the mean-square bending angle per unit length is the fundamental quantity which characterizes ``volumetric scattering'' sourced by a power spectrum of density fluctuations. The scattering time then scales linearly with the path length through the scattering material, multiplied by this mean-square bending angle. In addition, we have written concrete predictions for ``intermittent scattering'', sourced not by a power spectrum of density fluctuations within turbulent gas clumps, but, instead, by the bulk geometry of coherent filamentary or sheet-like structures with large aspect ratios. In both of these scenarios, we have shown that the distribution of FRB scattering timescales---or simply, the TDF---is sensitive to small-scale density fluctuations in the ionized CGM. In a power spectrum formalism, this scale can be as small as the ``inner scale'' which is difficult to resolve both observationally and even in modern hydrodynamical simulations. This work concretely demonstrates the potential to probe the CGM using deep, narrow-field FRB observations. Sensitive, narrow-field telescopes like MeerKAT, FAST, the DSA, and Murriyang, with repeated visits to single fields, could excel at providing dense grids of FRBs. For intervening halos at $\lesssim 100$ Mpc, the FRB backlights used need not be localized (as $d_{\rm eff}$ is effectively known \textit{a priori}). Widefield drift-scan telescopes like CHIME or BURSTT will also rapidly build up exposure time towards large solid angles on the sky.

\section{Acknowledgments}
We gratefully acknowledge Vicky Kaspi, Daniel Amouyal, Stella Ocker, Wenbin Lu, and the CHIME/FRB Collaboration for helpful comments and discussions.

\vspace{5mm}
\software{\texttt{numpy} \citep{harris2020array}, \texttt{matplotlib}~\citep{hunter2007matplotlib}
}

\bibliographystyle{aasjournal}
\bibliography{references}

\appendix 

\section{Scattering from a turbulent medium}
\label{app:turbulent_scattering}

For the sake of completeness, here we present a self-contained derivation of the volumetric scattering formulas developed elsewhere \citep{CordesRickett1998, cordes2016radio, Cordes2022redshift, ocker2025microphysics}. Consider a turbulent medium with electron density fluctuations given by
\begin{equation}
    P_{\delta n_e} (q) = C^2_n q^{-\beta} e^{-q^2 / q_i^2},\,\,(q_{\rm o} < q),
    \label{eq:power_spectrum}
\end{equation}
with outer and inner scales, $l_{\rm i} = 2\pi / q_{\rm i}$, $l_{\rm o} = 2\pi / q_{\rm o}$. Consider a slab of such a medium with thickness, $L$. We would like to compute the induced scattering time scale through the slab given the amplitude of the fluctuations and the inner and outer scale. 

\subsection{Mean ray deflection}

First, we want to derive the mean square scattering angle that a ray undergoes when encountering a turbulent medium per unit distance through the medium. The bending angle that a ray traveling in the $z$-direction undergoes is given by transverse gradients in the phase along the ray:
\begin{equation}
    {\bm \alpha} = \frac{\lambda}{2\pi}\nabla_\perp \phi.
\end{equation}
The phase along a ray through a slab of thickness, $s$, is given by
\begin{equation}
    \phi = \lambda r_e \int_0^s n_e({\bm x}_\perp, z) dz,
\end{equation}
where ${\bm x}_\perp$ is the co-ordinate in the plane perpendicular to the line of sight. We are assuming that the bending angles are small so that we can approximate the phase of the true ray as the phase computed along a straight line in the $z$ direction. 

We want to compute the statistics of this quantity by relating it to the power spectrum of the density fluctuations. Using the Fourier convention
\begin{align}
    \delta n_e({\bm x}) &= \int d^3 {\bm q} \delta \widetilde{n}_e({\bm q}) e^{i {\bm q} \cdot {\bm x}}, \\
    \langle \delta \widetilde{n}_e ({\bm q}) \delta \widetilde{n}^*_e({\bm q}') \rangle &= \delta^{(3)}({\bm q} - {\bm q}') P_{\delta n_e}(q),
\end{align}
we can write the bending angle through a slab of thickness, $s$, as
\begin{equation}
    {\bm \alpha} = \frac{\lambda^2 r_e}{2\pi} \int_0^s dz \int d^3 {\bm q} (i {\bm q}_\perp) \delta \widetilde{n}_e({\bm q}) e^{i ({\bm q}_\perp \cdot {\bm x}_\perp + q_z z)}.
\end{equation}
It follows that the mean square bending angle is
\begin{equation}
    \langle \alpha^2 \rangle = \frac{\lambda^4 r^2_e}{4 \pi^2} \int_0^s dz \int_0^s dz' \int d^3{\bm q} q^2_\perp P_{\delta n_e}(q) e^{i q_z (z - z'')}.
\end{equation}
Differentiating this with respect to the slab thickness, $s$, and applying the Leibniz integral rule, we obtain
\begin{equation}
    \frac{d \langle \alpha^2 \rangle}{ds} = \frac{\lambda^4 r^2_e}{4 \pi^2} \int_{-s}^s dt \int d^3{\bm q} q^2_\perp P_{\delta n_e}(q) e^{i q_z t}.
\end{equation}
Now, assuming the correlation lengths are small relative to the slab thickness, we let $s \to \infty$, so that the exponential in the integrand becomes a delta function, $2 \pi \delta(q_z)$. Replacing $q_{\perp} = q \sin \theta$ and $q_z = q \cos\theta$, we arrive at
\begin{align}
\begin{split}
    \frac{d \langle \alpha^2 \rangle}{ds} &= \frac{\lambda^4 r^2_e}{2 \pi} \int q^2 dq d\Omega q^2 \sin^2\theta P_{\delta n_e}(q) \delta(q \cos\theta), \\
    &= \lambda^4 r^2_e \int dq q^3 P_{\delta n_e}(q) \int d\theta \sin^3\theta \delta(\cos\theta).
\end{split}
\end{align}
The angular integral simply evaluates to unity. Following Cordes and Rickett, we call the mean deflection angle per unit length $\eta(s)$ and find
\begin{equation}
    \eta(s) = \lambda^4 r_e^2 \int dq q^3 P_{\delta n_e}(q, s),
    \label{eq:eta_s}
\end{equation}
where in this last line we note that the power spectrum can also vary along the ray, so long as it is does not vary faster than the outer scale of the turbulence. 

Note that \citet{CordesRickett1998} define $\eta(s)$ from the second moment of the angular brightness distribution, computed from the visibility function $\langle E({\bm x}) E^*({\bm x}+{\bm x_\perp)}\rangle = \exp(-0.5{D_\phi ({\bm x}_\perp}))$, where $D_\phi$ is the phase structure function \citep[see e.g.][]{Rickett1990}. In any case, one arrives at the same formula, Eq.~\ref{eq:eta_s}.

\subsection{Mean scattering time}
 
With the mean ray deflection angle in hand, we can compute the mean scattering time that a point source will exhibit under ray optics. For a power spectrum of the form Eq.~\ref{eq:power_spectrum}, one finds \citep{CordesRickett1998}:
\begin{equation}
    \eta(s) = \frac{\lambda^4 r^2_e \Gamma(3 - \beta/2)}{4 - \beta} q_{\rm i}^{4 - \beta} C^2_n(s),
\end{equation}
assuming that the change in the power spectrum along the line of sight is limited to changes in the amplitude of the fluctuations and that $\beta < 4$.

We would like to relate the amplitude $C^2_n$ to the (slightly) more observable quantities such as the mean electron density and the variance of the electron density. Consider a single cloud. The variance of the perturbed electron density for a scale $l$ is
\begin{equation}
    \sigma^2_{n_e} = \frac{2 (2\pi)^{4 - \beta} C^2_n l^{\beta-3}}{\beta - 3}
\end{equation}
for $\beta > 3 $ and $l \gg l_{\rm i}$. Thus, we can express the amplitude as
\begin{equation}
    C_n^2 = K_\beta \epsilon^2 n_{e,c}^2 l_{\rm o}^{3 -\beta},
\end{equation}
where $n_{e,c}$ is the local average electron density of the cloud and we have defined $\epsilon = \sigma_{n_{e,c}} / n_{e,c}$ to be the fractional rms variance of the electron density perturbation at the outer scale of the cloud. We have also defined the constant $K_\beta \equiv (\beta - 3) / 2 (2\pi)^{4 - \beta}$. 

Now let us consider a medium filled with an ensemble of such cloudlets, taking up a fraction of the total volume, $f_V$. In this case, we must compute the volume-averaged amplitude:
\begin{equation}
    \overline{C^2_n} = f_V K_\beta \langle \epsilon^2 n^2_{e, c} \rangle l^{3 - \beta}_{\rm o},
\end{equation}
where here the line, $\overline{\, \, \cdot \,\,}$, denotes a volume average, and the angle bracket, $\langle \cdot \rangle$, denotes an ensemble average. It is common to define an additional parameter, $\zeta \equiv \langle n^2_{e,c} \rangle / \langle n_{e ,c} \rangle^2$, which characterizes the fractional variation in the internal electron density of individual clouds in the ensemble. Thus, we can write
\begin{equation}
    \overline{C^2_n} = K_\beta \overline{n_e}^2 l_{\rm i}^{4 - \beta} \frac{\zeta \epsilon^2}{f_V l_{\rm o}^{\beta - 3} l_{\rm i}^{4 - \beta}} = K_\beta \overline{n_e}^2 l_{\rm i}^{4 - \beta} \widetilde{F},
\end{equation}
where $\overline{n_e} = f_V \langle n_{e,c} \rangle$ is the average electron density over the entire volume. We have also assumed that every cloud has the same fractional rms fluctuations so that $\langle \epsilon^2 n^2_{e,c} \rangle = \epsilon^2 \langle n^2_{e,c} \rangle$. Finally, the fluctuation parameter is defined:
\begin{equation}
    \widetilde{F} \equiv  \frac{\zeta \epsilon^2}{f_V l_{\rm o}^{\beta - 3} l_{\rm i}^{4 - \beta}},
    \label{eq:def_fl}
\end{equation}
and the mean squared deflection angle per unit length through the medium is
\begin{equation}
    \eta = A_\beta \lambda^4 r_e^2 \widetilde{F} \overline{n_e}^2,
    \label{eq:eta_F}
\end{equation}
where
\begin{equation}
    A_\beta = \frac{1}{2} \frac{\beta - 3}{4 - \beta} \Gamma(3 - \frac{\beta}{2}).
\end{equation}

The primary observable we are interested in is the scattering timescale. We can convert the mean square deflection per path length to the mean scattering time via:
\begin{equation}
    \tau_{\rm mean} = \frac{1}{2c} \int_0^{d_s} \eta(s) s (1 - \frac{s}{d_{so}}) ds.
\end{equation}
For a thin screen of thickness, $L$, and a roughly constant $\eta(s)$ throughout, one arrives at
\begin{align}
    \tau_{\rm mean} = \frac{d_{\rm eff}}{2c} L \eta.
\end{align}
where $d_{\rm eff} = d_{sl} d_{lo} / d_{so}$. In cosmological contexts, one must modify this to obtain
\begin{align}
    \tau_{\rm mean} = \frac{d_{\rm eff}}{2c} (1+z_l)^{-3} L \eta,
    \label{eq:tau_mean}
\end{align}
where all the distance now refer to angular diameter distances, $\lambda$ is the observing frequency, and $z_l$ is the redshift of the scattering medium. One obtains a factor of $(1+z_l)^{-4}$ as $\lambda \to (1+z_l)\lambda$ in the bending angle formula (Eq.~\ref{eq:eta_F}). An additional factor of $(1+z_l)$ is obtained, since $\tau$ is defined as the phase delay of the rays at the observer, $\tau = d\phi_O/d\omega_O$, which differs from the phase at the scatterer by a factor of $(1+z_l)$.

Together, Eqs.~\ref{eq:eta_F} and \ref{eq:tau_mean} provide the mean scattering time for an impulse scattered by a medium of thickness, $L$, filled with an ensemble of cloudlets making up a fraction of the total volume, $f_V$. We could, alternatively, consider a single cloudlet with $\widetilde{F}_c = f_V \widetilde{F}$. Then the scattering mean scattering angle through a single cloudlet of radius, $r_c$, is
\begin{align}
    &\tau_{\rm mean, c} = \frac{d_{\rm eff}}{2c} (1+z_l)^{-3}  r_c \eta_c, \\
    &\eta_c = A_\beta \lambda^4 r^2_e \widetilde{F}_l n^2_{e,c}.
\end{align}
Then, assuming that every cloudlet has the same fluctuation parameter, $\widetilde{F}_l$ and average density, $n_{e,c}$, the mean scattering time is simply given by
\begin{equation}
    \tau_{\rm mean} = \frac{d_{\rm eff}}{2c} (1 + z_l)^{-3} L_{\rm tot} \eta_c,
    \label{eq:tau_mean_singlecloud}
\end{equation}
where $L_{\rm tot}$ is the total path length for which a given line of sight intersects the cloudlets. If the cloudlets are confined to a medium of thickness $L$ and make up a fraction of the volume, $f_V$, then on average $L_{\rm tot} = f_V L$. With $\widetilde{F}_l = f_V \widetilde{F}$ and $n_{e,c} = \overline{n_e} / f_V$, one sees that Eq.~\ref{eq:tau_mean_singlecloud} is equivalent to Eq.~\ref{eq:tau_mean}. 

The advantage of this formulation is that it gives us an easy way of computing the statistics of the observed scattering times. The primary focus of this paper is computing the TDF, i.e. $f_{\rm sc}(>\tau)$. The key insight is that (assuming the cloudlets have uniform turbulent density fluctuation parameters), the variation in the observed scattering time is given precisely by the variation in the total path length intersecting the cloudlets, $L_{\rm tot}$. The statistics of this intersection length is sensitive to the cloudlet geometry.

One subtlety to mention is that the actual observed scattering time is the $1/e$ timescale seen in FRB scattering tails. This is related to the mean scattering time by
\begin{equation}
    \tau_{1/e} = A_\tau \tau_{\rm mean},
\end{equation}
where $A_\tau$ is a constant that depends on the thickness of the scattering medium and the relation between the inner scale and the diffractive scale \citep{lambert1999theory}. Since it does not change the shape of the scattering distributions, but shifts the mean by some constant offset, we will simply take $A_\tau = 1$.

\subsection{The role of diffractive scattering (wave optics)}
In the main text, we computed total scattering times using the mean deflection angle for rays through a turbulent medium under the assumption of ray (geometric) optics, neglecting diffractive contributions to the scattering time. Here we demonstrate that the diffractive contributions are negligible even when the inner scale of turbulent density fluctuations $l_i << r_\mathrm{diff}$.

Diffractive optics produces angular deflections of order $\theta_\mathrm{diff} = d_\mathrm{eff} r_\mathrm{diff}$, where $r_\mathrm{diff}$ is the transverse physical separation at which two rays will arrive at the observer with, on average, a radian offset in phase. For a medium with a power-law power spectrum, one finds~\citep{ocker2025microphysics}:
\begin{equation}
    r_{\rm diff} = \left[\beta \pi^2 2^{2-\beta}\frac{\Gamma(-\beta/2)}{\Gamma(\beta/2)} \lambda^2 r^2_e L \overline{C^2_n} \right]^{1/(2-\beta)}.
\end{equation}
Re-writing this in terms of the fluctuation parameters we have defined, one obtains
\begin{align}
\begin{split}
    r_{\rm diff} &= \left[ N_\beta r^2_e \lambda^2 L \overline{n_e}^2 l_{\rm i}^{4-\beta} \widetilde{F} \right]^{1/(2-\beta)}, \\
    N_\beta &= \beta (\beta - 3) 2^{\beta-3} \pi^{\beta-2} \frac{\Gamma(-\beta/2)}{\Gamma(\beta/2)}.
\end{split}
\end{align}
Now the typical wavefront deflection angle produced by diffractive scattering is $\alpha_d = \lambda / 2 \pi r_{\rm diff}$, which corresponds to a diffractive scattering time of
\begin{equation}
    \tau_d = \frac{d_{\rm eff}}{2c} \left( \frac{\lambda}{2\pi r_{\rm diff}}\right)^2
\end{equation}

We would like to compare this diffractive scattering time to the refractive (ray) scattering time of the previous section to determine which one predominates. To do so, we can re-write the diffractive scale in terms of the mean square deflection angle of the previous section, finding:
\begin{equation}
    r_{\rm diff} = \left[ \frac{N_\beta}{A_\beta} \lambda^{-2} l_{\rm i}^{4 - \beta} L \eta \right]^{1 / (2-\beta)}.
\end{equation}
Then
\begin{equation}
    \tau_d = \frac{d_{\rm eff}}{8\pi^2 c} \left( \frac{N_\beta}{A_\beta} \right)^{2/(\beta - 2)} (l_{\rm i} /\lambda)^{2(4-\beta)/(\beta-2)}(L\eta)^{2/(\beta - 2)}.
\end{equation}
Now, consider the quantity $r_{\rm diff} / l_{\rm i}$. This can be written as
\begin{equation}
    \frac{r_{\rm diff}}{l_{\rm i}} = \left( \frac{N_\beta}{A_\beta} \right)^{-1/(\beta-2)} (l_{\rm i} / \lambda)^{-2/(\beta-2)} (L\eta)^{-1/(\beta-2)}.
\end{equation}
Thus, whenever $r_{\rm diff} \gg l_{\rm i}$, we also have that
\begin{equation}
    (\lambda/l_i)^2 \gg \frac{N_\beta}{A_\beta} L\eta,
\end{equation}
from which it follows:
\begin{align}
\begin{split}
    \tau_d &\ll \frac{d_{\rm eff}}{8 \pi^2 c}  \left( \frac{N_\beta}{A_\beta} \right)^{2/(\beta - 2)} \left( \frac{N_\beta}{A_\beta} \right)^{-(4-\beta)/(\beta - 2)} (L\eta)^{-(4-\beta)/(\beta - 2)}(L\eta)^{2/(\beta - 2)}, \\
    &\ll \frac{d_{\rm eff}}{8 \pi^2 c}\frac{N_\beta}{A_\beta} L\eta, \\
    &\ll \frac{N_\beta}{4 \pi^2 A_\beta} \tau_{\rm mean},
\end{split}
\end{align}
assuming $2 < \beta < 4$. For Kolmogorov turbulence ($\beta = 11/3$), the constant $N_\beta / 4 \pi^2 A_\beta \approx 2.8$. Thus, when the inner scale is much less than the diffractive scale (or, equivalently, when $\lambda/l_{\rm i} \gg \sqrt{L \eta}$), the refractive mean scattering time exceeds the diffractive scattering time. For simplicity, we will assume this is the case, and only apply the ray optics formulae for the scattering time. However, we note that, whereas the mean refractive scattering time scales exactly linearly with the effective intersection length, for a Kolmogorov turbulent medium, the diffractive scattering time scales nearly linearly: $\tau_d \propto L^{6/5}$. Thus, we argue that qualitatively, our results will be similar for the $r_{\rm diff} \ll l_i$ regime. The regime where diffractive optics transitions to refractive optics, $l_i \sim r_{\rm diff}$, is not well characterized and may require more advanced wave optics techniques \cite{}. 

\section{A multi-scale mist}
\label{app:multi-scale}

In Section~\ref{sec:volumetric_scattering}, we compute the statistics of the intersection length $L$ for a complex of spherical cloudlets with uniform radius. Here we consider a spectrum of cloudlet sizes. The resulting distributions are the same, with the exception that the cloudlet optical depth must be modified. 

Consider spherical cloudlets with a power-law distribution of radii,
\begin{equation}
    P(r_c) \propto r_c^{-a},
\end{equation}
between some minimum and maximum sizes, $r_{c,\rm min}$ and $r_{c,\rm max}$. Let $f_V$ be the volume filling factor of the cloudlets in a region of volume $V$. The total number of cloudlets in that region is
\begin{align}
\begin{split}
    n_V &=  f_V V / \int_{r_{c,\rm min}}^{r_{c,\rm max}} P(r_c) \frac{4}{3} \pi r_c^3 dr_c,\\
    &= f_V V \frac{3}{4\pi} \frac{4 - a}{1 - a} \frac{z_{c,\rm max}^{1-a} - z_{c,\rm min}^{1-a}}{z_{c,\rm max}^{4-a} - z_{c,\rm min}^{4-a}}.
\end{split}
\end{align}
The average intersection length for lines of sight through the gas complex is given by
\begin{align}
\begin{split}
    \overline{L} &= \int_{r_{\rm c, min}}^{r_{\rm c, max}} r_c \frac{\pi r_c^2}{A} n_V P(r_c) dr_c,
\end{split}
\end{align}
where $A$ is the corresponding sky area of the total volume. Evaluating the integral, one obtains $\overline{L} = f_V R$, as expected. 

\begin{figure}
    \centering
    \includegraphics[width=\columnwidth]{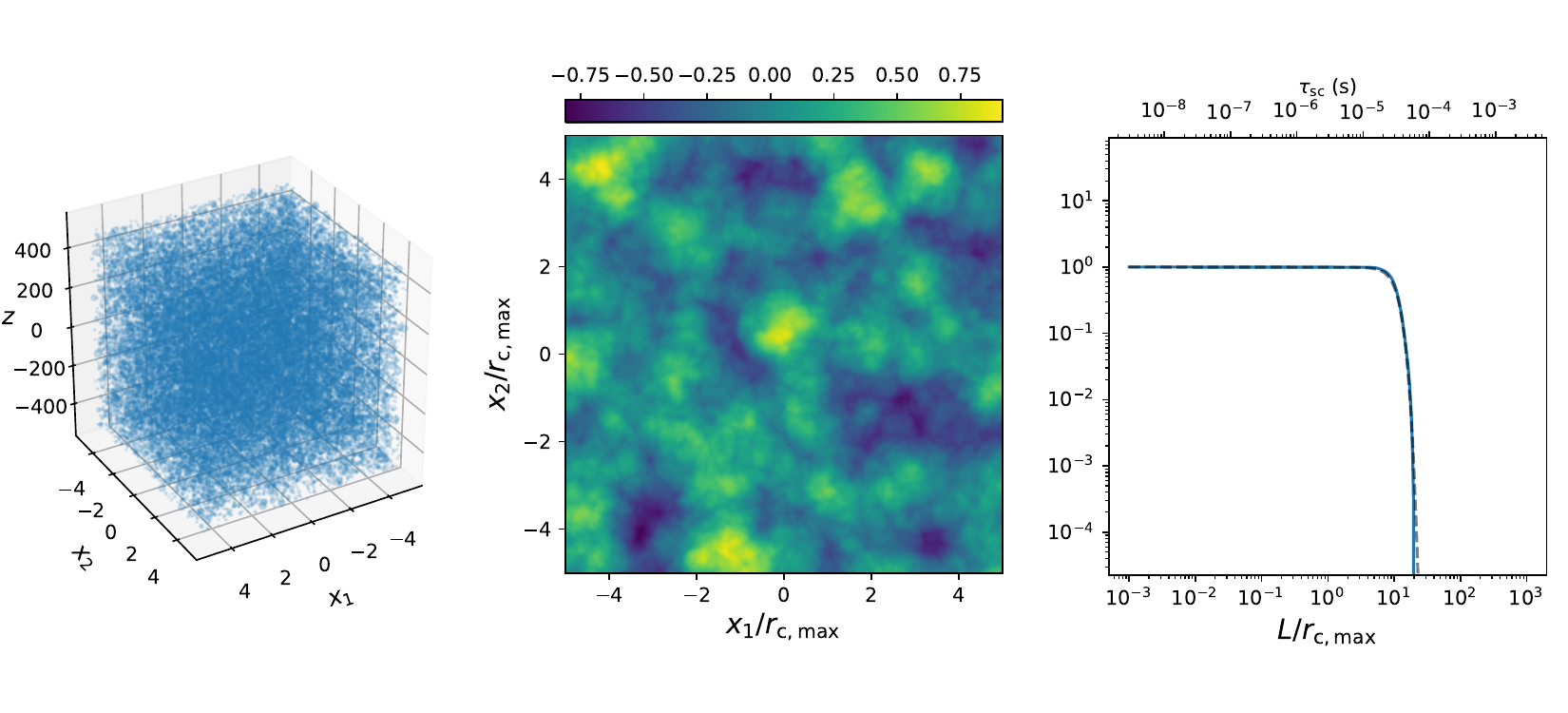}
    \caption{A simulation of intersection lengths through a complex of spherical cloudlets with a power-law distribution of radii ($\alpha = 2$, $r_{\rm c, min} = 0.1\,{\rm pc}$, $r_{\rm c, max} = 10\,{\rm pc}$). The left panel shows the mist of cloudlets, randomly distributed in a box of dimensions $10 r_{\rm c, max} \times 10 r_{\rm c, max} \times R$, with $R = 1\,{\rm kpc}$. The co-ordinate axes are in units of $r_{\rm c, max}$. The middle panel shows the intersection length contrast, $\delta L = (L - \overline{L}) / \overline{L}$, for a grid of vertical sight-lines through the box, and the right shows the cumulative distribution function of these intersection lengths in blue. The dashed line shows the analytic curve for a Gaussian distribution (Eq.~\ref{eq:ftau_sphere_analytic}).}
    \label{fig:multi-scale-mist}
\end{figure}

Now, we will define an effective cloudlet optical depth to be the minimum number of intersected clouds:
\begin{equation}
    N_c = f_V R / r_{\rm c,max}.
\end{equation}
So long as $N_c > 1$, the largest cloudlets dominate the variance of the intersection length distribution, and one finds that $\delta L \sim \mathcal{N}(0, N_c^{-1})$, as with the single-scale mist. It follows that the form of the TDF is identical to Eq.~\ref{eq:ftau_sphere_analytic}, except that the effective optical depth, $N_c$, is modified to be referenced to the largest cloud size (see Fig.~\ref{fig:multi-scale-mist}).

\section{Scattering from a refractive slab}
\label{app:slab}

In this section, we compute the characteristic bending angle for a slab with electron density $n_e$, thickness $\delta$, and length $\Lambda$, as a function of its inclination angle $\iota$ with respect to the line of sight. Since the bending angle a ray undergoes when encountering free electrons is simply determined by the transverse gradient in the excess electron column, we need to estimate the excess column density of the slab. As can be seen from Fig.~\ref{fig:sheet_diagram}, the column density of a thin slab is given by $N_e = n_e l_\parallel = n_e \delta / \cos\iota$. Thus, a face-on sheet ($\iota = 0$) has a column density of $n_e \delta$. If the slab were infinitely long, an edge-on slab ($\iota = \pi  /2$) would have an infinite column density as $\cos\iota \to 0$. However, for a physical slab with finite length $\Lambda$, we will take the column density to be simply
\begin{equation}
    N_e (\iota) = \min\left\{ \frac{n_e\delta}{\cos\iota}, n_e\Lambda \right\}.
\end{equation}
To estimate the transverse gradient of the column density, we need to know the transverse scale over which the column density varies. This is set by $l_\perp = \min\left\{\delta / \sin\iota, \Lambda\right\}$. We can thereby estimate the transverse column density gradient as $\nabla_\perp N_e \sim n_e l_\parallel / l_\perp$. From this, we obtain the maximum bending angle that such a slab can produce is
\begin{equation}
    \alpha_{\rm max}(\iota) = \frac{\lambda^2 r_e}{2\pi} \nabla_\perp N_e \sim \frac{\lambda^2 r_e n_e}{2 \pi} \min\left\{ \tan \iota, A \right\},
\end{equation}
where $A = \Lambda / \delta$ is the aspect ratio of the sheet. The bending angle increases with inclination until it saturates to some maximum value determined by $A$. There is also a minimum value that is reached as $\iota \to 0$, due to the largest possible value of $l_\perp = \Lambda$; however, since in this work we are concerned with the cumulative distribution function of scattering timescales above some value, we will simply neglect this minimum value.  Now, for a single slab, we will simply take the characteristic bending angle to be half the maximum possible bending angle, defining:
\begin{equation}
    \alpha_{\rm slab}(\iota) = \alpha_0
    \begin{cases}
        \tan \iota, & \,\,\, \iota < {\rm arctan}{A}, \\
        A, & \,\,\, \iota > {\rm arctan}A,
    \end{cases}
\label{eq:alpha_slab}
\end{equation}
where $\alpha_0 = \lambda^2 r_e n_e / 4 \pi$.

In order to predict the TDF for a complex of such slabs, we need to know how many slabs, on average, scatter light into the line of sight with a scattering delay $> \tau_*$. Consider an annulus in the scattering plane with angular radii $[\alpha_*, \alpha_{\rm max}]$, where $\tau_* = d_{lo} \alpha_*^2/2c$ and $\alpha_{\rm max} = \alpha_0 A$ is the maximum bending angle that such slabs can produce\footnote{Here, we assume that the source is at infinity so that $d_{\rm eff} = d_{\rm lo}$ and the bending angle $\alpha$ corresponds to the observing angle}. We can divide this annulus into smaller, concentric annuli $[\alpha, \alpha + d\alpha]$.

Now, $N_{\rm slab} = f_V R/\Lambda^2 \delta = f_V R / A^2 \delta^3$ is the number column density for sight-lines through a complex of size $R$. Then, an annulus with bounds $[\alpha, \alpha+d\alpha]$ has, on average, 
\begin{equation}
    dn(\alpha) = N_{\rm slab} \cdot 2 \pi (d_{lo} \alpha) (d_{lo} d\alpha) \cdot p(\iota)
\end{equation}
sheets that scatter light into the line of sight. We need to weight the number of sheets in the annulus by the probability that it has an inclination angle $\iota$ such that the bending angle is $\alpha = \alpha_{\rm slab}(\iota)$. For uniform random inclinations, the probability distribution is $p(\iota) = \sin\iota$.

Assuming a Poisson process, the probability that there are no sheets within this small annulus which scatter light into the line of sight is $P_{\rm none}(\alpha) = e^{-dn(\alpha)}$. The probability that \textit{none} of the annuli between $[\alpha_{\rm min}, \alpha_{\rm max}]$ contain slabs which scatter light into the line of sight is $P(<\tau_{\rm min}) = \Pi_{\alpha_*}^{\alpha_{\rm max}} P_{\rm none}(\alpha)$. In the limit $d\alpha \to 0$, this is:
\begin{equation}
    P(<\tau_*) = \exp\left\{- 2 \pi N_{\rm slab} d_{lo}^2 \int_{\alpha_*}^{\alpha_{\rm max}} \alpha(\iota) \sin\iota d\alpha \right\}.
\end{equation}
We can re-write the integral over bending angel as an integral over all inclinations from $[\iota_*, \pi/2]$, where $\iota_*$ is given by $\alpha_* = \alpha(\iota_*)$. We must split this integral into two parts: $[\iota_*, \iota_{\rm max}]$, where $\iota_{\rm max} = \arctan A$ is the inclination angle where $\alpha_{\rm max} = \alpha_0 A$ is achieved, and $[\iota_{\rm max}, \pi/2]$. In the former region, using $d\alpha = \alpha_0 \sec^2\iota d\iota$, we obtain
\begin{align}
\begin{split}
    P(<\tau_*) &= \exp
    \left\{- 2 \pi N_{\rm slab} d_{lo}^2 \alpha_0^2 
    \left[
    \int_{\iota_*}^{\iota_{\rm max}} \tan\iota \sin\iota \sec^2\iota d\iota
    + \int_{\iota_{\rm max}}^{\pi/2} A^2 \sin\iota d\iota 
    \right]
    \right\} \\
    &= \exp\left\{- 2 \pi N_{\rm slab} d_{lo}^2 \alpha_0^2 
    \left( \left[\frac{1}{2} \tan\iota \sec\iota - \frac{1}{2}\arctan(\sin\iota)\right]^{\iota_{\rm max} = \arctan A}_{\iota_* =\arctan \alpha_*/\alpha_0} + A^2 \cos\iota_{\rm max}\right)
    \right\} \\
     &= \exp\left\{-\pi N_{\rm slab} d_{lo}^2 \alpha_0^2
    \left( \left[ x \sqrt{x^2 + 1} - \arctan \left( \frac{x}{\sqrt{x^2+1}}\right) \right]^{x=A}_{x=\alpha_*/\alpha_0} + \frac{A^2}{\sqrt{A^2 + 1}}\right)
    \right\}.
\end{split}
\end{align}
Finally,
\begin{align}
\begin{split}
    P(>\tau_*) &= 1 - P(< \tau_*) \\
    &= 1 - \exp\left\{-\frac{f_V R}{A^2 \delta^3} d_{lo}^2 \frac{n_e^2 r_e^2 \lambda^4}{16 \pi}
    \left( \left[ x \sqrt{x^2 + 1} - \arctan \left( \frac{x}{\sqrt{x^2+1}}\right) \right]^{x=A}_{x=\alpha_*/\alpha_0} + \frac{A^2}{\sqrt{A^2 + 1}}\right)
    \right\}
    \label{eq:TDF_slab}
\end{split}
\end{align}
Note that for $\tau_* = 0$ (and for $A \gg 1$) this is simply
\begin{equation}
    P(>0) = 1 - \exp\left\{ - \pi n R D_l^2 \alpha_0^2 A^2 \right\} = 1 - \exp \left\{ - N_{\rm slab} \pi d_{lo}^2 \alpha^2_{\rm max} \right\}.
\end{equation}
The quantity $N_{\rm slab} \pi d_{lo}^2 \alpha_{\rm max}^2$ is just the total number of sheets within the maximum scattering distance, as expected. 

We also wish to compute the average number of slabs intersected by a given line of sight, $N_c$, in order to compare with the volumetric scattering regime. For a complex of size $R$ and a volume filling factor, $f_V$, of slabs, the average path length intersecting such slabs is $f_V R$. To compute the average number of slabs contributing to this path length, we need to know the average path length through a single slab integrated over arbitrary inclination angles. For a slab with large aspect ratio ($A \gg 1$), we can split this integral into two parts. When the slab is inclined by $\iota > \frac{\pi}{2} - A^{-1}$, we will consider this to be edge on, so that the path length along the line of sight through the slab is simply given by the total length, $\Lambda$. When the slab is inclined by $\iota < \frac{\pi}{2} - A^{-1}$, the intersection length is given by $\delta / \cos\iota$. We want to average this over the solid angle $d\Omega = \sin \iota d\iota d\phi$. Thus, we obtain the average intersection length through a single slab to be:
\begin{align}
\begin{split}
    \overline{l}_\parallel&= \int_{\iota =0}^{\iota =\pi/2 - A^{-1}} \frac{\delta}{\cos \iota} d\Omega + \int_{\iota = \pi/2 - A^{-1}}^{\iota = \pi/2} \Lambda d\Omega, \\
    &= -2 \pi \delta \log \left( \sin A^{-1}\right) + 2\pi \Lambda \sin A^{-1}, \\
    &\approx 2\pi \delta \left(1 + \log A\right),
\end{split}
\end{align}
where the last line is obtained for $A \gg 1$. It follows that the average number of slabs intersected by a given line of sight is $N_c = f_VR / \overline{l}_\parallel$:

\begin{figure}
    \centering
    \includegraphics[width=0.5\linewidth]{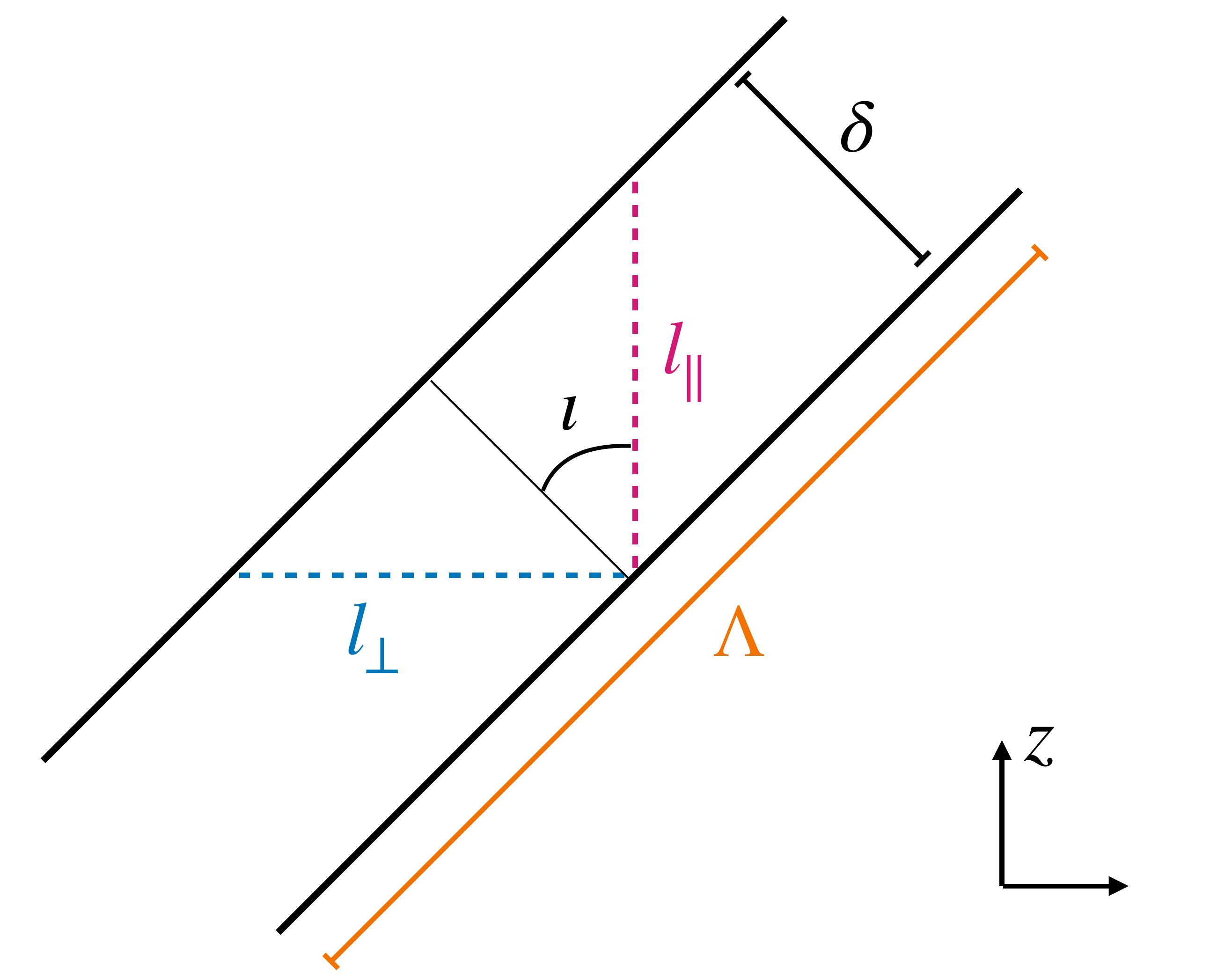}
    \caption{Diagram of an inclined slab of scattering material with length $\Lambda$ and width $\delta$. The vertical $z$-axis is the line of sight and $\iota$ is the inclination angle. The lengths $l_\parallel$ and $l_\perp$ define length-scales of the slab projected parallel and perpendicular to the line of sight, respectively.}
    \label{fig:sheet_diagram}
\end{figure}

\subsection{Note on multiple refractive images}

We have now used Eq.~\ref{eq:alpha_slab} to compute the TDF. Eq.~\ref{eq:alpha_slab} gives the bending angle of a slab as a function of its inclination, which corresponds to a scattering time. Assuming a uniform distribution of inclination angles, we can compute  the statistics of the scattering time. While this approach allows for a simple expression for the TDF, there are a few subtleties, which we will discuss here. Firstly, as we discuss in the case of intermittent, refractive scattering from uniform spherical clouds---in order to produce multi-path propagation (which then may observed as a scattering timescale), the lens must have sufficiently large convergence ($\kappa \gtrsim 1$). The convergence of slabs is enhanced by a factor of $A$ relative to a spherical geometry, and thus anisotropic structures are likely to be more relevant for intermittent scattering scenarios. In Section~\ref{sec:intermittent}, we compute the TDF assuming a sufficiently large $\kappa$ to produce scattering; however, when using observations to place constraints on the underlying parameters, case must be taken. Null observations will have no constraining power on regions of parameter space with $\kappa \ll 1$. Secondly, while scattering timescales are typically identified by measuring the temporal broadening of radio pulses, intermittent, refractive scattering may not necessarily produce characteristic scattering tails. Indeed, a single slab will only produce order-unity additional images. This few-image regime is often referred to as ``plasma lensing"~\citep{1998ApJ...496..253C, CordesRickett1998,cordes2017lensing} instead of scattering (though the underlying physics is the same), and is more challenging to identify in the dynamic spectra of pulsars and FRBs due to the many parameters in the modeling of individual plasma lenses (see e.g.~\citealp{kader2025detection}). Many slabs with random orientations may also only produce a handful of images. If the slabs are organized into a larger structure, such that they inherit similar inclination angles, then many images may be produced, recovering the scattering-tail phenomenology. A similar scenario has been proposed for pulsar scintillation. \citet{pen2014pulsar} argue that refractive current sheets in the ISM, sustained by magnetic fields, can produce hundreds of images. Small transverse perturbations propagate along the sheet as shallow waves, forming lasagna-like corrugations. Each individual corrugation can be treated, in a simple approximation, as a single slab. A corrugated sheet, therefore, can be viewed as a series of off-set slabs aligned along the length of the current sheet.

\end{document}